\documentclass{emulateapj}

\usepackage{graphicx}
\usepackage{amsmath}
\usepackage{natbib}
\usepackage{amssymb}

\usepackage[breaklinks,colorlinks,urlcolor=blue,citecolor=blue,linkcolor=blue]{hyperref}
\usepackage{hyperref}

\newcommand{\be}{\begin{eqnarray}}
\newcommand{\ee}{\end{eqnarray}}

\newcommand{\lp}{\left(}
\newcommand{\rp}{\right)}
\newcommand{\lb}{\left[}
\newcommand{\rb}{\right]}

\newcommand{\slugcom}{Accepted for publication in The Astronomical Journal}
\slugcomment{\slugcom}
\begin{document}

\normalsize


\title{Can Rocky Exoplanets with Rings Pose as Sub-Neptunes?}

\author{Anthony L. Piro}

\affil{The Observatories of the Carnegie Institution for Science, 813 Santa Barbara St., Pasadena, CA 91101, USA; piro@carnegiescience.edu}

\begin{abstract}
In our solar system, the presence of rings is exclusive to the gas giants, but is this the case for all planetary systems? In principle, it seems that rocky exoplanets could also have rings, which could be searched for by studying their subtle imprint on the ingress and egress of transits. Unfortunately, such effects are difficult to measure and require high precision photometric and/or spectroscopic observations. At the most basic level though, the presence of rings would result in an increased transit depth that could be mistaken as an anonymously large radius. Motivated by this, I consider how a population of exoplanets with rings would impact radius measurements, focusing on Earth-like exoplanets. It is found that this population introduces an enhancement of inferred radii in the range of $\sim2-3R_\oplus$, not unlike the sub-Neptunes that have been identified in recent transit surveys. Whether rings can explain all or most sub-Neptunes seems difficult, since it would require a large fraction of rocky planets to have rings ($\gtrsim40\%$) and/or a factor of $\sim2-3$ increase in the number of planets with radii $\lesssim1.2R_\oplus$. Even if rings cannot explain all sub-Neptunes, this work suggests that focusing on those planets currently classified as sub-Neptunes may be a good starting place for finding rocky planets with rings.
\end{abstract}


\section{Introduction}
\label{sec:introduction}

Rings are common amongst the four outer gas-rich planets in out solar system, while they are completely absent for the rocky planets {(although rings are present for some of the smaller rocky bodies, \citealp{Braga14,Ortiz15,Ortiz17}).} Nevertheless, if we have learned anything from the last two decades of exoplanet research, it is that other planetary systems find a way of repeatedly defying our expectations based on what we know from closer to home. This thus begs the question whether the terrestrial extrasolar planets, which are now known to be basically ubiquitous in our Galaxy \citep{Burke15,Mulders15}, actually sometimes have rings.

In fact, there are reasons to think that some rocky planets will have rings. Phobos is currently migrating inward toward Mars on a relatively short timescale of $\sim70\,{\rm Myr}$, and it will likely tidally disrupt and form rings in the future \citep{Black15}. It has been suggested that this process has occurred repeatedly throughout Mar's history, so that Mars has alternatively had a moon or rings on timescales of $\sim100\,{\rm Myr}$ \citep{Hesselbrock17}. In other theoretical work, it has been suggested that the combined tides of a moon and parent star on an exoplanet can in some cases send the moon migrating into the planet until it is tidally disrupted and forms rings \citep{Counselman73,Barnes02,Sasaki12,Piro18}. Thus, determining the presence or absence of rings may be an important probe of a planet's  history.

In principle, rings should be detectable from detailed photometric or spectroscopic changes to transits that are strongest during ingress and egress \citep[e.g.,][]{Barnes04,Ohta09,Zuluaga15} or from variations from transit to transit \citep{Sucerquia17}. The difficulty is that such signals are subtle and can be difficult to discern in current data. In a few cases, potential rings or at least constraints on rings have been made in this way \citep{Heising15,Aizawa17}, and in at least one case it has been argued that an exoplanet has a giant ring system from a series of complex eclipses \citep{Kenworthy15,Rieder16}, but this work has largely focused on giant planets rather than the rocky planets discussed here.

Ignoring these details, the simplest impact of rings would be to increase the depth of transits such that instead of measuring the planet radius $R_p$, an eclipsed area of $A$ will result in an inferred radius of
\be
	R_{\rm inf} = (A/\pi)^{1/2}\gtrsim R_p.
    \label{eq:r_inf}
\ee
Thus if a population of exoplanets are found with seemingly anomalous radii, for example, because the exoplanets have densities that are too low \citep{Zuluaga15}, this may indicate we are observing $R_{\rm inf}$ rather than $R_p$. Potentially connected with this is the issue of exoplanets with radii larger than Earth's radius $R_\oplus$, but smaller than Neptune \citep{Batalha13}. There are no direct analogs of such planets in our solar system, which has led to interest in their origin. At least for the planets with radii $\gtrsim1.6\,R_\oplus$, the so-called sub-Neptunes, radial velocity measurements \citep{Marcy14} and detailed transit timing modeling \citep{Wu13,Hadden14,Hadden17} reveal that most of these planets have low densities inconsistent with a purely rocky composition. The likely explanation is that they have extended, gaseous envelopes \citep{Weiss14,Rogers15}, but given that rings could also increase the radius without increasing the mass appreciably, it seems like this is another option that should be considered.

Motivated by these issues, here I explore the inferred radii of rocky planets if there is a population with rings. In Section~\ref{sec:lifetime}, I discuss the properties of these rings, focusing on their expected size and lifetime. In Section~\ref{sec:area}, I derive the eclipsed area of an exoplanet with rings as a function of the ring properties and viewing angle. In Section~\ref{sec:montecarlo}, I perform Monte Carlo calculations to understand the distribution of inferred radii for a population of ringed exoplanets, which are them compared to observations of super-Earths and sub-Neptunes in Section~\ref{sec:comparison}. Finally, I conclude in Section~\ref{sec:conclusion} with a summary of this work and a discussion of future areas to investigate.

\section{Size and Lifetime of a Rocky Ring}
\label{sec:lifetime}

Following a scenario that may generate rings around a rocky planet \citep[e.g.,][]{Hesselbrock17,Piro18}, the ring material will spread due to viscous effects. This can be the result of a combination of self-gravity wakes and local collisional processes \citep{Salmon10}. Here I address whether rings should last for a sufficiently long time to be observationally important.

\begin{figure}
\begin{center}
  \includegraphics[width=0.45\textwidth,trim=0.4cm 0.3cm 0.4cm 0.2cm, clip]{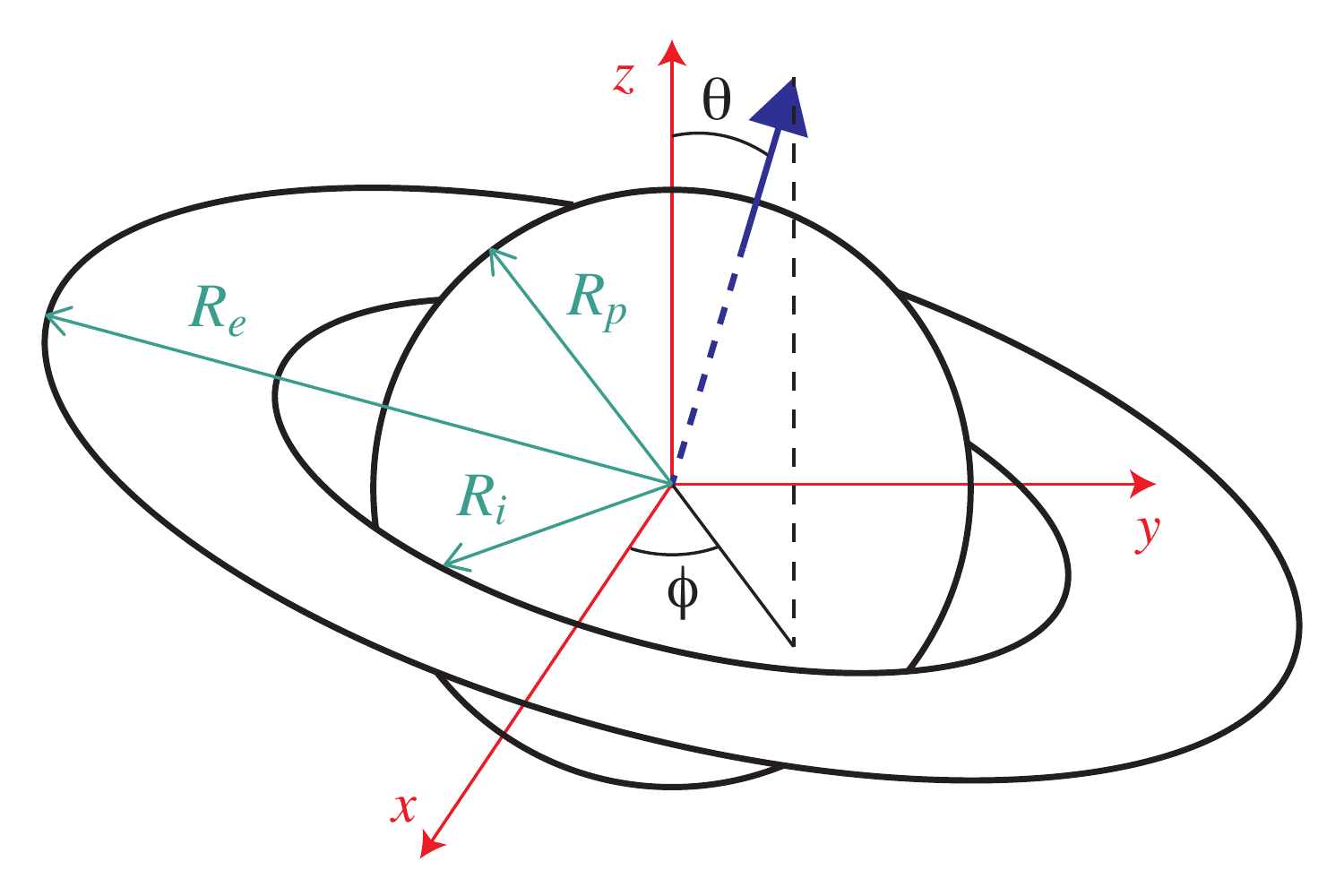}
    \end{center}
  \caption{Planet with radius $R_p$ encircled by rings with inner and out edge radii $R_i$ and $R_e$, respectively. It is assumed that $R_e\approx a_{\rm FRL}$, the fluid Roche limit, and that $R_i\approx R_e/2$, similar to Saturn's rings. For an observer sitting along the $x$-axis, the tilt of the rings can be described by two angles, the obliquity with respect to the $z$-axis, $\theta$, and the azimuthal twist, $\phi$. The orientation of the rings is represented by the normal vector shown by the thick blue arrow. The area seen by the observer is simply this normal vector projected onto the $x$-axis, thus the effective area of just the rings is $\pi (R_e^2-R_i^2)\sin\theta|\cos\phi|$.}
  \label{fig:geometry}
\end{figure}

Rings as pictured in Figure \ref{fig:geometry} will viscously spread out until the outer edge $R_e$ reaches the fluid Roche limit \citep{Murray99}
\be
	R_e = a_{\rm FRL} \approx 2.46 R_p \lp\frac{\rho_p}{\rho} \rp^{1/3},
    \label{eq:rd}
\ee
where $\rho_p$ is the bulk density of the planet ($\rho_p\approx5.5\,{\rm g\,cm^{-3}}$ for the Earth) and $\rho$ is the density of particles that make up the rings. This density should not be confused with the rings themselves but rather the individual particles. Since the rings are envisioned as being rocky, a characteristic density of $\rho\approx 2\,{\rm g\,cm^{-3}}$ will be used, motivated by the type of material that makes up Phobos. Outside of the radius $a_{\rm FRL}$, material will aggregate into satellites and no longer be part of the ring.

A key point is that the rings evolve in a self-similar way with time. This means that the evolution is rather quick at early times, and its final state is not too sensitive to the initial conditions \citep[for example, see the ring evolution solutions of][]{Salmon10}. Thus $R_e$ quickly reaches $a_{\rm FRL}$, and this can be approximated as the outer ring radius for the majority of its lifetime. Assuming that the densities of planets and ring material do not vary strongly, then $R_e$ is simply proportional to the radius of the planet (and anyway, the density dependence is weakened because of the cube root). In addition, there is the inner radius of the rings $R_i$, the exact value of which is less certain than $R_e$. Similar to the rings of Saturn, I will approximate $R_i\approx R_e/2$, although the exact value for $R_i$ does not qualitatively change any of the conclusions in this work.

This self-similar evolution is also important for understanding the lifetime over which the rings are expected to be optically thick, since the total lifetime will be dominated by the viscous timescale at the latest stages. For rings with a surface density $\Sigma$ composed of particles with radius $r$ and density $\rho$, the optical depth is
\be
	\tau \approx \Sigma /r\rho.
\ee
Thus if the rings are marginally optically thick with $\tau\approx1$, then $\Sigma \approx r\rho$, and the total mass of the rings from inner radius $R_i$ to outer edge $R_e$ is
\be
	M_r &\approx& \pi \Sigma (R_e^2-R_i^2) \approx 3\pi r\rho R_e^2/4
    \nonumber
    \\
	&\approx & 2.3\times10^{21}
      	r_{100}\rho_2^{1/3}
    	\lp\frac{M_p}{M_\oplus}\rp^{2/3}{\rm g},
\ee
where $r_{100}=r/100\,{\rm cm}$, $\rho_2=\rho/2\,{\rm g\,cm^{-3}}$, and $M_\oplus$ is the mass of Earth. As a comparison, this is about two orders of magnitude more mass than Phobos, about $5\%$ of the mass of Saturn's rings, or about $0.2\%$ of the mass of Ceres.

The nature of the viscous evolution changes greatly depending on whether the ring is self-gravitating or not. This is measured by the Toomre $Q$ parameter \citep{Toomre64}
\be
	Q = \frac{\Omega\sigma_r}{3.36 G\Sigma},
\ee
where $\Omega=(GM_p/R^3)^{1/2}$ is the orbital frequency at a radius $R$ and $\sigma_r$ is the particle radial velocity dispersion. Roughly speaking, a ring will be gravitationally unstable for $Q\lesssim 1$, although even for $Q\lesssim 2$ N-body simulations show that gravitational wakes can be effective at transporting angular momentum \citep{Salo95}. The velocity dispersion is regulated to roughly be the particle's escape velocity,
\be
	\sigma_r \approx (4\pi G \rho/3)^{1/2}r \approx 7.5\times10^{-2} r_{100}\rho_2^{1/2}\,{\rm cm\,s^{-1}},
    \label{eq:sigma_r}
\ee
and the characteristic orbital frequency at $R$ is
\be
	\Omega \approx 1.9\times10^{-4}
    \rho_2^{1/2} \lp \frac{R}{R_e}\rp^{-3/2}{\rm s^{-1}}.
\ee
Putting these together,
\be
	Q\approx 0.32 \lp \frac{R}{R_e}\rp^{-3/2},
\ee
independent of the composition of the ring. Thus an optically thick rocky ring is gravitational unstable and self-gravity wakes will provide large scale transportation of angular momentum. Using the results of \citet{Daisaka01}, the associated self-gravity viscosity can be approximated as
\be
	\nu_{\rm SG} = 13 \lp \frac{r_{\rm H}}{2r}\rp^5
    \lp \frac{G^2\Sigma^2}{\Omega^3}\rp,
\ee
where
\be
	r_{\rm H}=\lp \frac{8\pi \rho r^3}{9M_p}\rp^{1/3} R_e
    \approx 2.1 r\lp \frac{R}{R_e}\rp,
\ee
is the particle's Hill radius. The fact that $r_{\rm H}/r\gtrsim 1$ justifies the use of the particle's escape velocity for $\sigma_r$ in Equation~(\ref{eq:sigma_r}), otherwise, if $r_{\rm H}/r\lesssim 1$, then the velocity dispersion would be dominated by the relative Keplerian velocity between particles $\approx 2r\Omega$ \citep{Daisaka01}. The viscosity is then
\be
	\nu_{\rm SG} &=& 430 r_{100}^2\rho_2^{1/2}\lp \frac{R}{R_e}\rp^{19/2}
    {\rm cm^2\,s^{-1}}.
\ee
Another possibility is that angular momentum can be transported via sound waves traveling between colliding particles, which has an associated viscosity \citep{Araki86,Wisdom88}
\be
	\nu_{\rm col} = r^2\Omega\tau
    =1.9 r_{100}^2\rho_2^{1/2}
    \lp \frac{R}{R_e}\rp^{-3/2}{\rm cm^2\,s^{-1}},
\ee
where again $\tau\approx1$ is assumed. Since $\nu_{\rm col}\lesssim \nu_{\rm SG}$, the viscous timescale is dominated by self-gravity effects. Thus, the timescale for the rings to viscously evolve is estimated as
\be
	t_{\rm visc} &\approx& R^2/\nu_{\rm SG}
    \nonumber
    \\
    &\approx& 3.6\times10^8 r_{100}^{-2}\rho_2^{-7/6}
    \lp\frac{M_p}{M_\oplus}\rp^{2/3}
    \lp \frac{R}{R_e}\rp^{-15/2}
    {\rm yr}.
    \nonumber
    \\
\ee
This demonstrates that once formed, these rings are expected to be optically thick for a significant amount of time. This is similar to the timescales found in more detailed simulations for hypothesized rings around Mars during its previous history \citep{Hesselbrock17}.

\section{Eclipsed Area of a Ringed Exoplanet}
\label{sec:area}

Next, I consider the area an exoplanet with rings eclipses from the point of view of an observer watching a transit. As shown in Figure~\ref{fig:geometry}, the parent star, exoplanet, and observer are assumed to approximately sit in the same $xy$-plane since eclipses only occur for a small range of separation angles $\approx R_*/D\ll 1$, where $R_*$ is the star's radius and $D$ is the distance between the star and exoplanet. The orientation of the ring with respect to the observer depends on two angles. These are the obliquity $\theta$, which is measured from the $z$-axis perpendicular to the exoplanet's orbit, and the azimuthal angle $\phi$ (also sometimes referred to as the ``season''). With these definitions, the values of these angles run from $0\le \theta \le \theta_{\rm obl}$, where $\theta_{\rm obl}$ is the maximum possible obliquity for these systems, and $0\le \phi \le 2\pi$.

\begin{figure}
\begin{center}
  \includegraphics[width=0.44\textwidth,trim=0.0cm 1.5cm 0.0cm 1.5cm, clip]{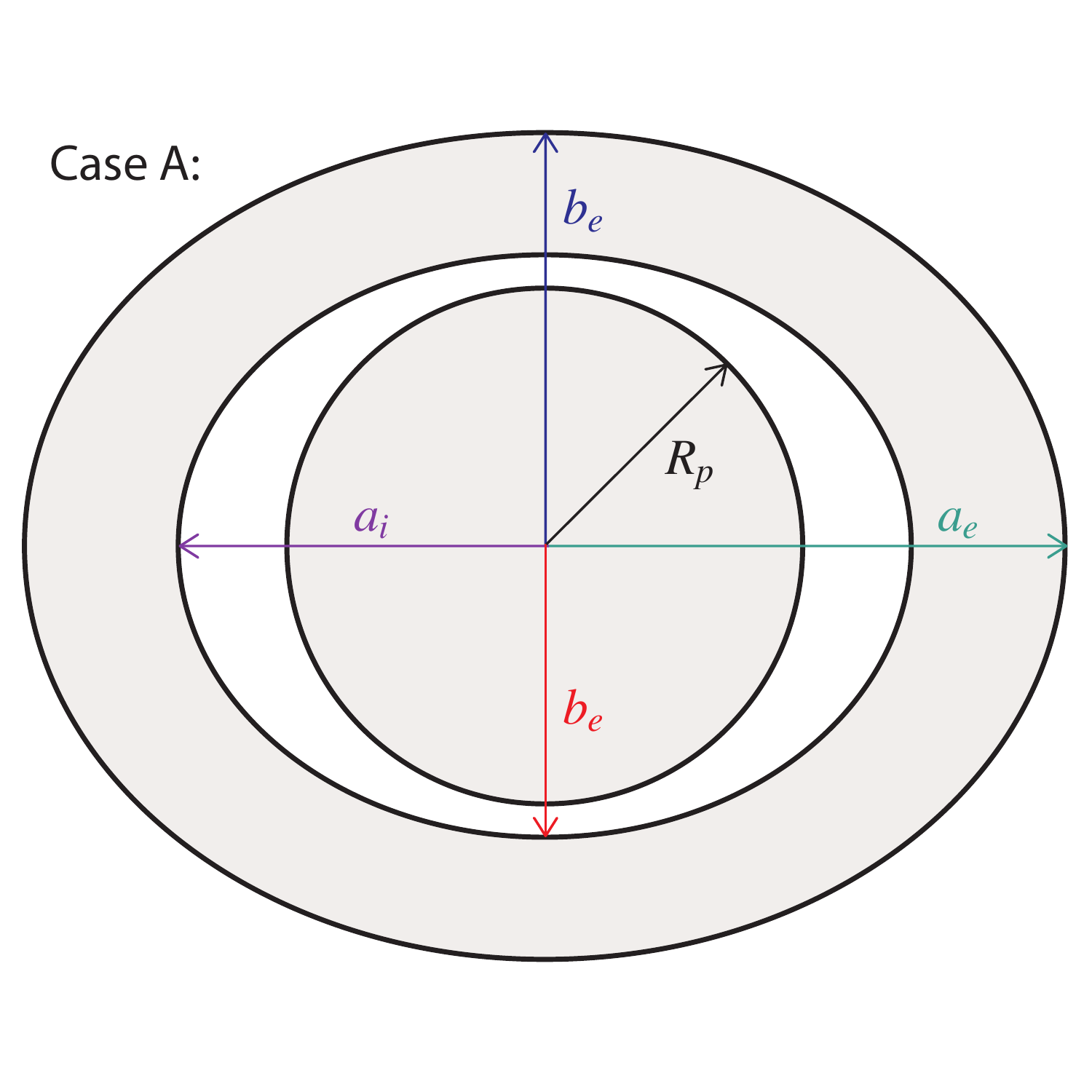}
    \includegraphics[width=0.44\textwidth,trim=0.0cm 3.0cm 0.0cm 2.0cm, clip]{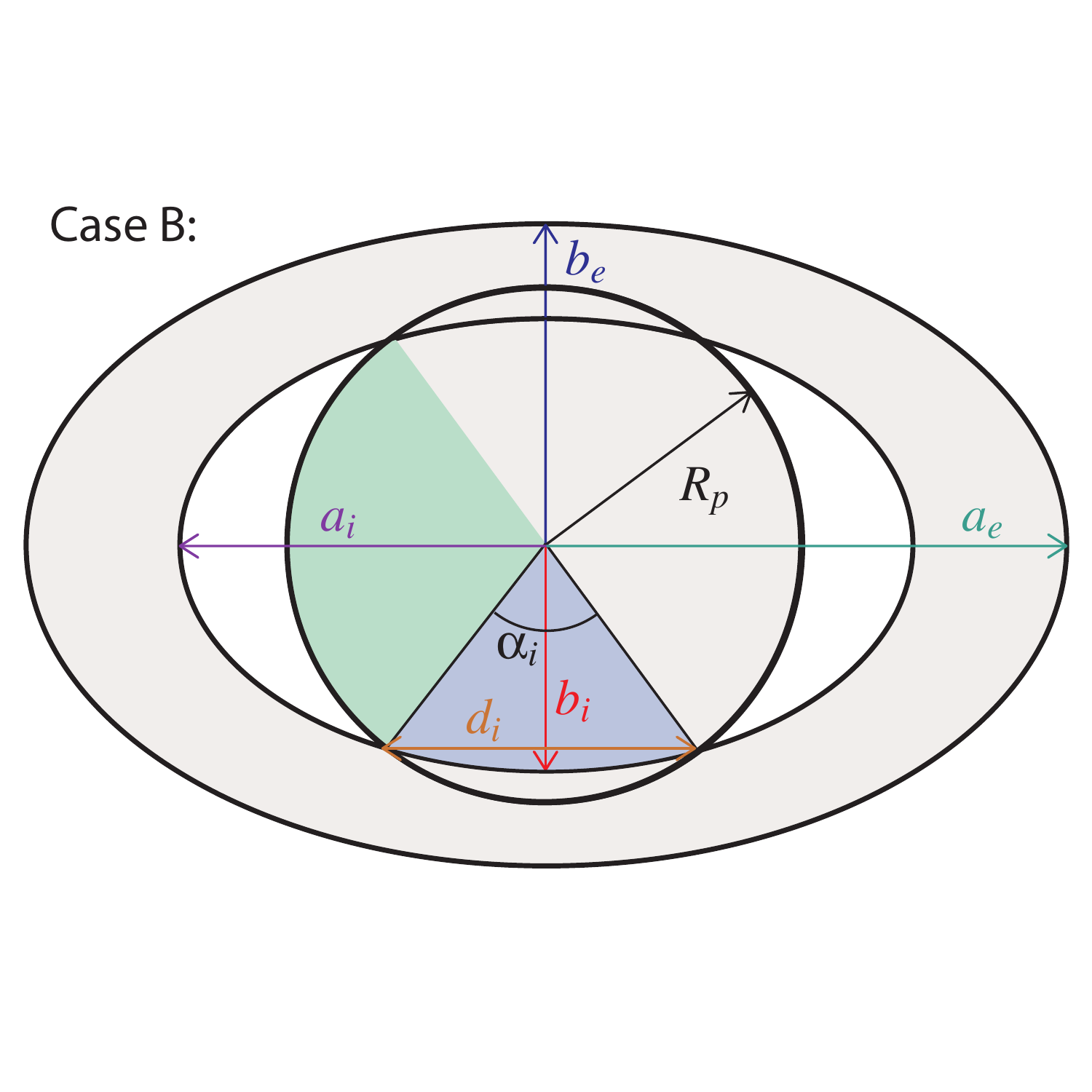}
    \includegraphics[width=0.44\textwidth,trim=0.0cm 3.5cm 0.0cm 3.0cm, clip]{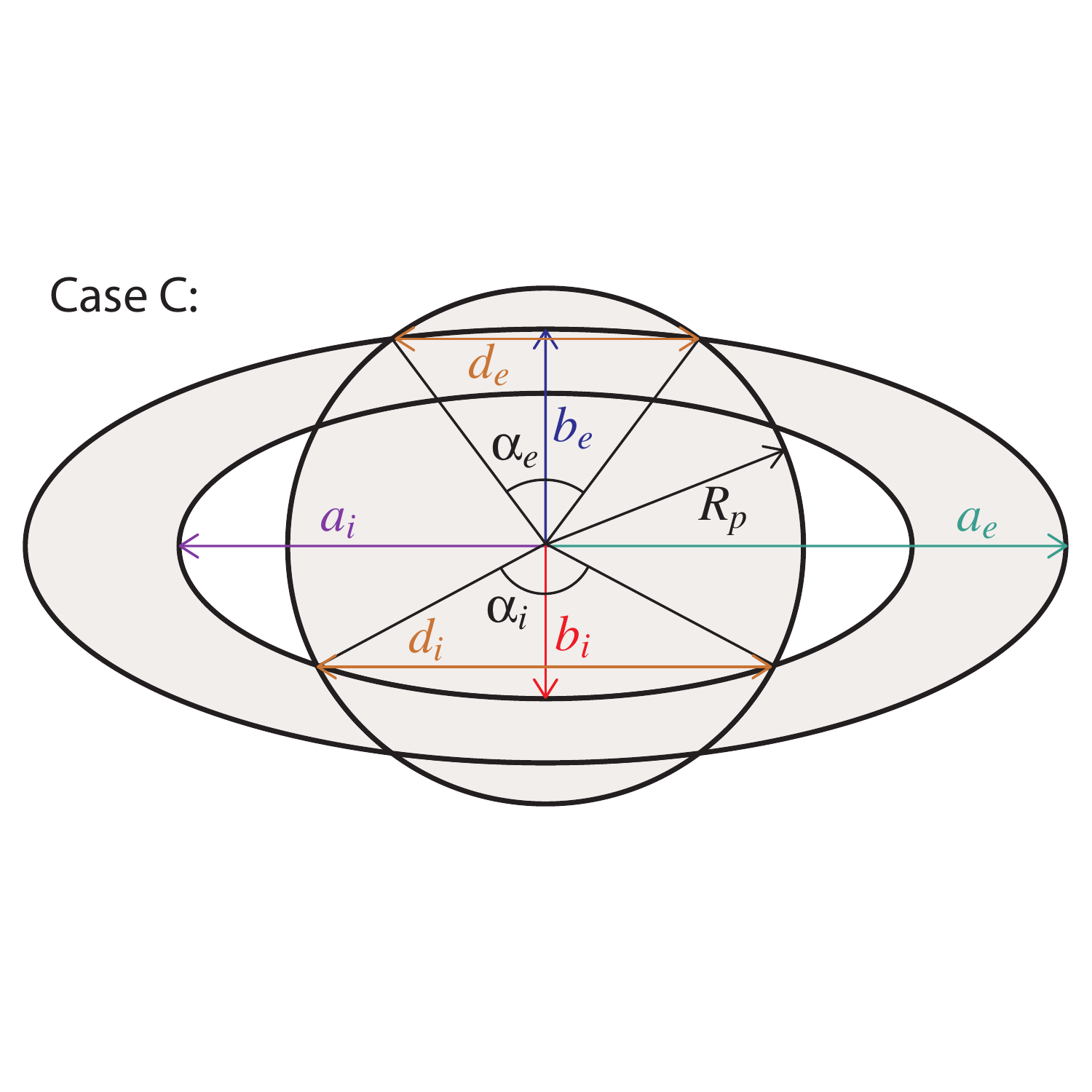}
 \end{center}
  \caption{Diagrams of the area covered by a planet plus ring. The ring has inner and outer edge radii $R_i$ and $R_e$, respectively. When viewed with obliquity $\theta$ from an angle $\phi$, these form ellipses with semi-major axes $a_i=R_i$ and $a_e=R_e$, respectively, and semi-minor axes $b_i=R_i\sin\theta|\cos\phi|$ and $b_e=R_e\sin\theta|\cos\phi|$, respectively. To simplify assessing the area, three separate cases are considered, Case A when $b_i>R_p$, Case B when $b_e>R_p>b_i$, and Case C when $R_p>b_e$. Other key angles and distances labeled in the panels are further described in the text.}
  \label{fig:area}
\end{figure}

To an observer, the inner and outer edge of the ring appear as an ellipses with semi-major axes $a_i=R_i$ and $a_e=R_e$, respectively, and semi-minor axes $b_i=R_i\sin\theta|\cos\phi|$ and $b_e=R_e\sin\theta|\cos\phi|$, respectively. The total area blocked by the ring is thus $\pi (a_eb_e-a_ib_i)=\pi(R_e^2-R_i^2)\sin\theta|\cos\phi|$, while the exoplanet will block out an area $\pi R_p^2$. This does not take into account that there are regions where the ring and planet overlap from the observer's point of view. To address this, three different cases are considered (see Figure \ref{fig:area}).

\subsection{Case A: $b_i>R_p$}

As shown in the top diagram of Figure~\ref{fig:area}, this is the simplest case where there is no overlap between the planet and the ring. Here the blocked area is
\be
	A = \pi(R_e^2-R_i^2)\sin\theta|\cos\phi| + \pi R_p^2,
\ee
which is simply the sum of the two areas.

\subsection{Case B: $b_e>R_p>b_i$}

As shown in the middle diagram of Figure~\ref{fig:area}, in this case the planet will cross over the inner edge of the ring, but does not extend beyond the ring's outer edge. First, there is the area covered by the ring, which just like before has a projected area of $\pi(R_e^2-R_i^2)\sin\theta|\cos\phi|$. Then, there is the remaining area covered by the planet. This can be divided into four sectors, two of which are highlighted in light blue and green. The area of the light blue sector can be found using Cavallieri's principle to be
\be
	{\rm Blue\; sector} = \frac{\alpha_i}{2}a_ib_i = \frac{\alpha_i}{2}R_i^2\sin\theta|\cos\phi|,
\ee
while the green sector is just a fraction of a circle
\be
	{\rm Green\; sector} = \frac{\pi-\alpha_i}{2}R_p^2.
\ee
The angle $\alpha_i$ can be found using some trigonometry. First note that the distance $d_i$ is given by
\be
	d_i = 2a_i\lp \frac{R_p^2-b_i^2}{a_i^2-b_i^2} \rp^{1/2}.
\ee
Thus the angle is
\be
	\alpha_i = 2 \arcsin
    	\lb \frac{a_i}{R_p}\lp \frac{R_p^2-b_i^2}{a_i^2-b_i^2} \rp^{1/2} \rb.
    \label{eq:alpha}
\ee
Putting this all together,
\be
	A &=& \pi (a_eb_e-a_ib_i) + \alpha_ia_ib_i +(\pi-\alpha_i)R_p^2,
    \nonumber
    \\
    &=& \pi (R_e^2-R_i^2)\sin\theta|\cos\phi| + \alpha_i R_i\sin\theta|\cos\phi|
    \nonumber
    \\
    && + (\pi-\alpha_i)R_p^2.
\ee
is the total observed area.

\subsection{Case C: $R_p>b_e$}

In the final case, the planet extends further than the outer edge of the ring from the observer's view. If the inner radius of the ring is ignored, one can find the area of the outer outline in the bottom panel of Figure \ref{fig:area}. This is composed of two elliptical sectors (on the left and right) and two circular sectors (on the top and bottom). Again using Cavallieri's principle,
\be
	{\rm Area\;of\;outer\;shape} = (\pi-\alpha_e)a_eb_e + \alpha_e R_p^2,
\ee
where in analogy with the derivation of $a_i$ above, one can first find the length
\be
	d_e = 2a_e\lp \frac{R_p^2-b_e^2}{a_e^2-b_e^2} \rp^{1/2},
\ee
to find the angle
\be
	\alpha_e = 2 \arcsin
    	\lb \frac{a_e}{R_p}\lp \frac{R_p^2-b_e^2}{a_e^2-b_e^2} \rp^{1/2} \rb.
    \label{eq:alpha_e}
\ee
The area of the inner ring then needs to be subtracted. This is done by subtracting the total elliptical sectors on the left and right and then adding back the area covered by the planet
\be
	{\rm Area\;of\;inner\;holes} = (\pi-\alpha_i)a_ib_i -(\pi-\alpha_i)R_p^2,
\ee
where $\alpha_i$ is the same as given by Equation (\ref{eq:alpha}) above. Putting this all together,
\be
	A &=& (\pi -\alpha_e)a_e b_e
    - (\pi-\alpha_i)a_i b_i + (\pi+\alpha_e-\alpha_i)R_p^2,
    \nonumber
    \\
    &=& (\pi -\alpha_e)R_e^2\sin\theta|\cos\phi|
    - (\pi-\alpha_i)R_i^2\sin\theta|\cos\phi|
    \nonumber
    \\
    &&+ (\pi+\alpha_e-\alpha_i)R_p^2,
\ee
is the total projected area.

\begin{figure}
\epsscale{1.0}
  \includegraphics[width=0.48\textwidth,trim=0.0cm 0.0cm 0.2cm 0.5cm, clip]{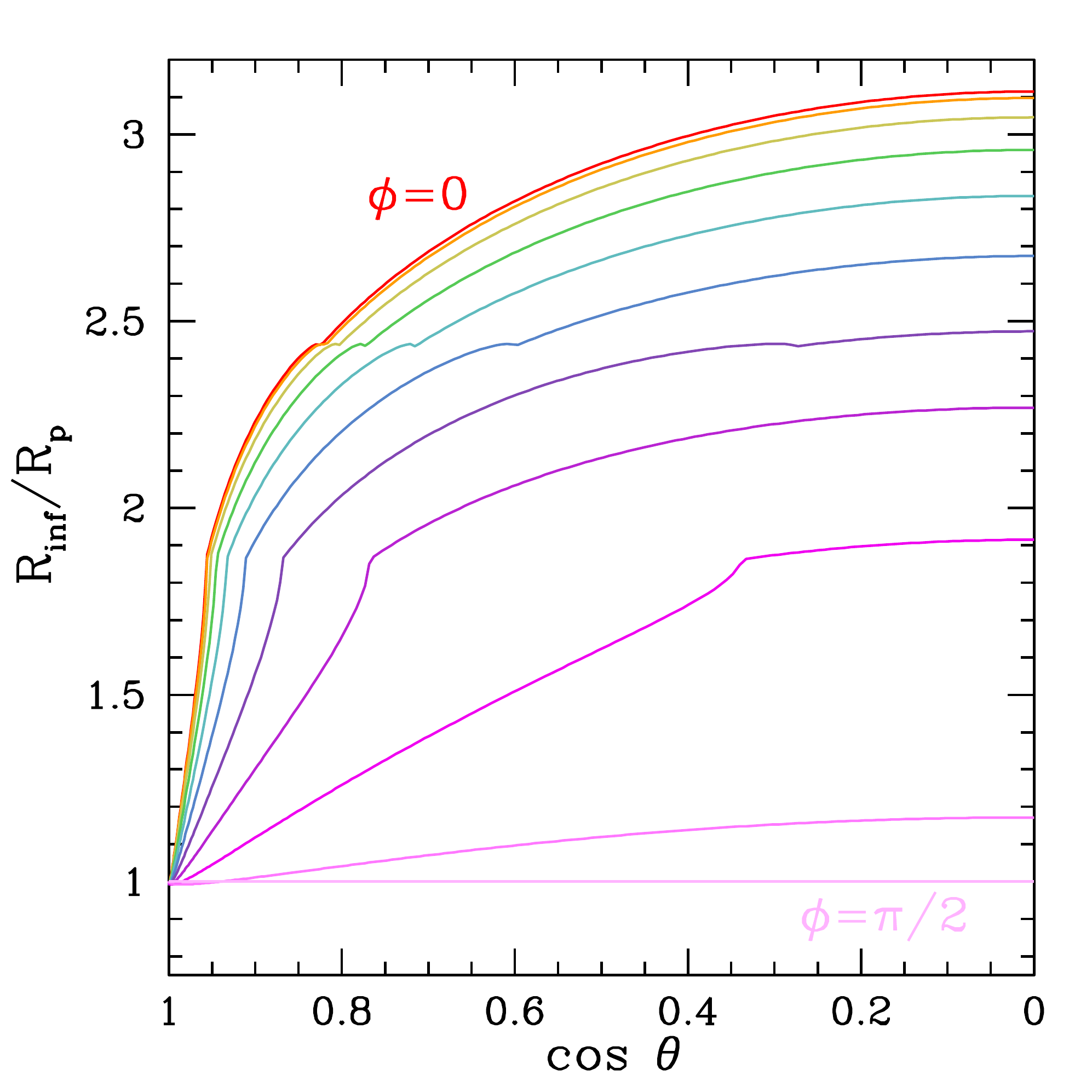}
\caption{Inferred radius $R_{\rm inf}$ for a planet with rings viewed at different angles $\theta$. Colors curves go from $\phi=0$ (red) to $\phi=\pi/2$ (light pink) in increments of $\pi/20$. The rings are chosen to have inner and outer edge radii of $R_i=1.72R_p$ and $R_e=3.44R_p$, respectively.}
\label{fig:radius}
\epsscale{1.0}
\end{figure}

\subsection{Inferred Radius}

If a transit is observed and the eclipse depth provides an area $A$ for the planet, then the inferred radius of the planet is given by Equation (\ref{eq:r_inf}). As shown for the values of $A$ derived above for a planet with rings, it is always the case that $R_{\inf}>R_p$. This is further shown in Figure~\ref{fig:radius}, where $R_{\rm inf}$ is plotted for different values of $\theta$ and $\phi$ (also see Figure 1 from \citealp{Heising15}, which shows example schematics for how the rings appear from different angles). At small $\theta$ (large $\cos\theta$), the ring is viewed nearly edge on and thus $R_{\rm inf}/R_p\approx 1$. As $\theta$ increases ($\cos\theta$ decreases), so does $R_{\rm inf}$ until reaching a maximum value when face on. Although $R_{\inf}$ changes continuously, kinks are seen as $R_{\rm inf}$ transitions though the three cases, starting at Case C (when $\theta$ is small) and ending with Case A (when $\theta$ is large, although note that not all three cases are covered when $\phi$ is sufficiently large).

These results assume that the rings are optically thick to the star light, but this assumption is justified for two reasons. First, the timescale discussion in Section~\ref{sec:lifetime} shows that it is plausible for the rings to remain optically thick for $\gtrsim10^9\,{\rm yrs}$. Second, the rings will in general be viewed from an angle different than exactly face on. The optical depth through the rings viewed at an angle is greater because the thickness along the observer's path is larger, so that it is more likely that the rings will appear optically thick.

\section{Monte Carlo Calculations}
\label{sec:montecarlo}

Although the impact of rings on a single transit may be subtle, a large collection of transits will show a component of anomalously large radii from a population of rings. Since there is a large range of parameters and distributions for the planetary properties, here I explore different possibilities using Monte Carlo calculations and summarize the impact on the distributions of $R_{\rm inf}$.

\begin{figure}
\epsscale{1.0}
  \includegraphics[width=0.48\textwidth,trim=0.0cm 0.0cm 0.2cm 0.5cm, clip]{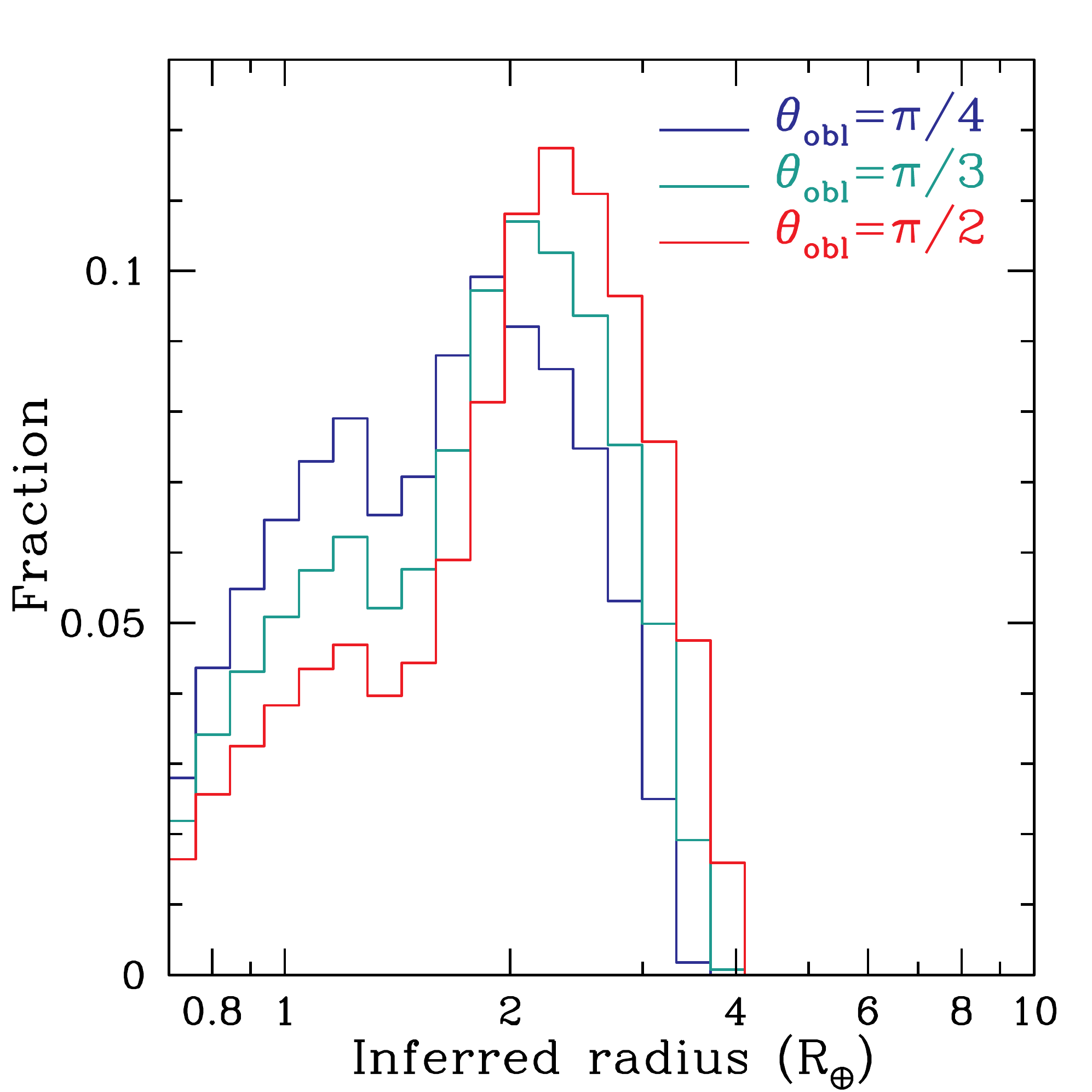}
\caption{Example inferred radius distributions for a population of rocky planets with rings for different maximum obliquities $\theta_{\rm obl}$. This assumes a power law distribution for the planetary radii $P(R_p)\propto R_p^{-1}$ with a range of $0.7\le R_p/R_\oplus \le 1.3$, along with flat distributions in $0\le \phi\le 2\pi$ and $\cos\theta_{\rm obl}\le \cos\theta \le 1$. The inner and outer edge of the rings have radii $R_i=1.72R_p$ and $R_e=3.44R_p$, respectively, which is appropriate for bulk densities for the planet and rings of $\rho_p=5.5\,{\rm g\,cm^{-3}}$ and $\rho=2\,{\rm g\,cm^{-3}}$, respectively. All planets are assumed to have rings.}
\label{fig:histogram}
\epsscale{1.0}
\end{figure}

\begin{figure}
\epsscale{1.0}
  \includegraphics[width=0.48\textwidth,trim=0.0cm 0.0cm 0.2cm 0.5cm, clip]{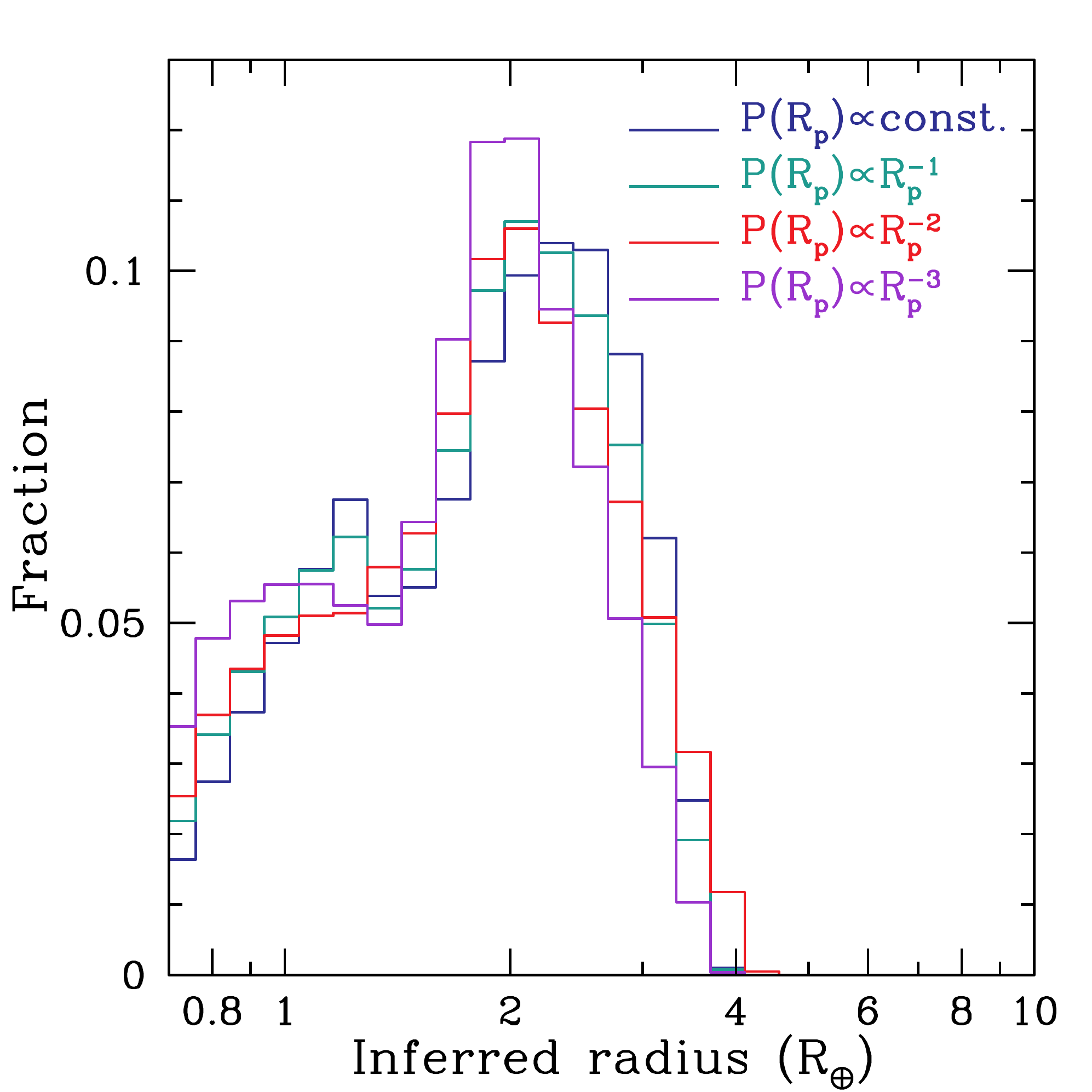}
\caption{Similar to Figure \ref{fig:histogram}, but with a fixed value of $\theta_{\rm obl}=\pi/3$ and different distributions for the exoplanet radii (see text and Appendix for a full description of these distributions).}
\label{fig:histogram2}
\epsscale{1.0}
\end{figure}

\begin{figure}
\epsscale{1.0}
  \includegraphics[width=0.48\textwidth,trim=0.0cm 0.0cm 0.2cm 0.5cm, clip]{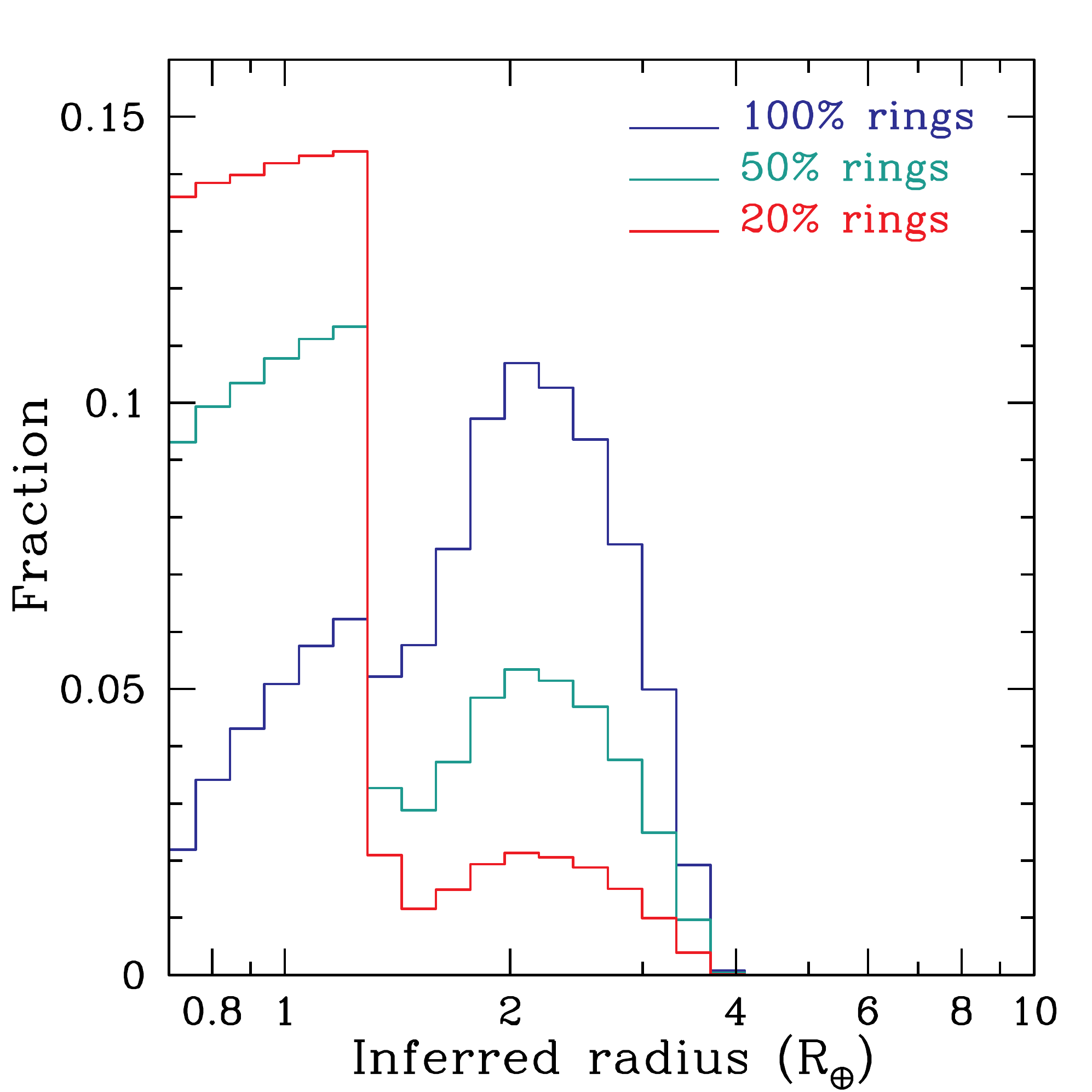}
\caption{Similar to Figure \ref{fig:histogram}, but with a fixed value of $\theta_{\rm obl}=\pi/3$ and $P(R_p)\propto R_p^{-1}$ with different fractions of planets with rings as labeled.}
\label{fig:histogram3}
\epsscale{1.0}
\end{figure}

The general strategy is to start with a probability density for the planetary radii
\be
	P(R_p) \propto R_p^{-\eta},
    \label{eq:probability}
\ee
which is defined such that
\be
	\int_{R_1}^{R_2} P(R_p) dR_p = 1,
\ee
where $R_1$ and $R_2$ are the minimum and maximum radii of rocky exoplanets. These are very uncertain, and although $R_1=0.7R_\oplus$ and $R_2=1.3R_\oplus$ are chosen for this work, this should be updated as this distribution is better understood. See the Appendix for further details on how $P(R_p)$ is implemented.

Using Equation (\ref{eq:rd}), the corresponding outer ring radius $R_e$ is estimated from a given $R_p$. The inner edge is less clear and for this work it is estimated as $R_i\approx R_e/2$, which is similar to Saturn's rings. For the angles, flat distributions in $0\le \phi\le 2\pi$ and $\cos\theta_{\rm obl}\le \cos\theta \le 1$ are considered. The transiting area is calculated using the three cases outlined in Section~\ref{sec:area}, and then I find $R_{\rm inf}$ using Equation~(\ref{eq:r_inf}). Typically $5\times10^6$ exoplanets are run to make sure the distributions converge. The histograms presented are made with logarithmically spaced bins, motivated by the work of \citet{Fulton17} on the radius distribution of small planets from California-{\it Kepler} Survey (further discussed in Section \ref{sec:comparison}). Such binning puts additional emphasis on the larger radii to highlight additional features that might be less apparent with more evenly distributed binning. 

A first example of a histogram of inferred radii is shown in Figure \ref{fig:histogram}. Here it is assumed that all exoplanets have rings and that $P(R_p)\propto R_p^{-1}$. The maximum obliquity $\theta_{\rm obl}$ is varied to better understand how this impacts the distribution of inferred radii. The obliquity of the planets in our solar system vary greatly, so three characteristic values are considered as denoted in Figure \ref{fig:histogram}. These examples show that a bimodal distribution of radii is generally expected when rings are present. This is because there is one larger radius peak from the average viewing angle of the rings, but in addition, there is a smaller radius peak from the nearly edge-on exoplanets where the transit is dominated by the exoplanet radius itself. The large radius peak increases in size and average value of $R_{\rm inf}$ with increasing $\theta_{\rm obl}$ because this results in more systems with rings closer to face on.

In Figure \ref{fig:histogram2}, I explore how the assumed planet distribution changes the bimodal distribution due the presence of rings. The main change is that a flatter distribution of $R_p$ results in a slightly larger value for the peak $R_{\rm inf}$, although this effect is more subtle that changing $\theta_{\rm obl}$ (see Figure \ref{fig:histogram}). At the low radius end, the radius distribution can be studied more directly, so that in the future hopefully this distribution can be constrained for input into studies such as the one here.

In the previous example, all planets are assumed to have rings, but this is clearly too optimistic. In Figure \ref{fig:histogram2}, I explore how the distributions change when only some fraction of planets have rings. Here, $\theta_{\rm obl}=\pi/3$ and $P(R_p)\propto R_p^{-1}$ and three different fractions of planets with rings are considered. Not surprisingly, the lower radius peak increases dramatically when the number of planets without rings is increased. This means that it is unlikely that the two peak in the radius distribution are comparable, unless there is a significant number of planets with rings.

\section{Comparison to Small Planet Observations}
\label{sec:comparison}

As mentioned in Section \ref{sec:introduction}, one of the developing areas in exoplanet studies is the large number of planets smaller than Neptune discovered by the {\it Kepler} mission. These were not predicted by initial theories of planet formation \citep[e.g.,][]{Ida04,Mordasini09}, which expected that planets should either fail to accrete enough material to become super-Earths or grow quickly and efficiently to form gas-rich giants. More recent formation models are now able to produce super-Earths as observed \citep[e.g.,][]{Hansen12,Mordasini12,Alibert13,Chiang13,Chatterjee14,Coleman14,Lee14,Raymond14,Lee16}, and in particular, now highlight the important role atmospheric erosion by photoevaporation can play in the planet radii distribution \citep[e.g.,][]{Owen13,Jin14,Lopez14,Chen16,Lopez16,Owen17}. Such models predict a dearth of intermediate sub-Neptune planets in highly irradiated environments. This is because a small envelope of H/He can greatly inflate a planet's radius, so that the result is either a bare rocky cores with a small radius or a rocky core with a small envelope but much larger radius \citep{Lopez14}. It was therefore interesting to see, using the subsample of planets from the California-{\it Kepler} Survey \citep{Johnson17,Petigura17,Petigura18,Weiss18}, a deficit in the occurrence rate distribution at a radius of $\approx1.5-2.0\,R_\oplus$ \citet{Fulton17}, somewhat consistent with this picture (although see \citealp{Bouma18} and \citealp{Teske18} for discussions on uncertainties in this distribution, depending on the presence of unseen companion stars). {This was also seen for a subsample of exoplanets that have parent stars with highly accurate stellar parameters determined from asteroseismology \citep{VanEylen17}.}

\begin{figure}
\epsscale{1.0}
  \includegraphics[width=0.48\textwidth,trim=0.0cm 0.0cm 0.2cm 0.5cm, clip]{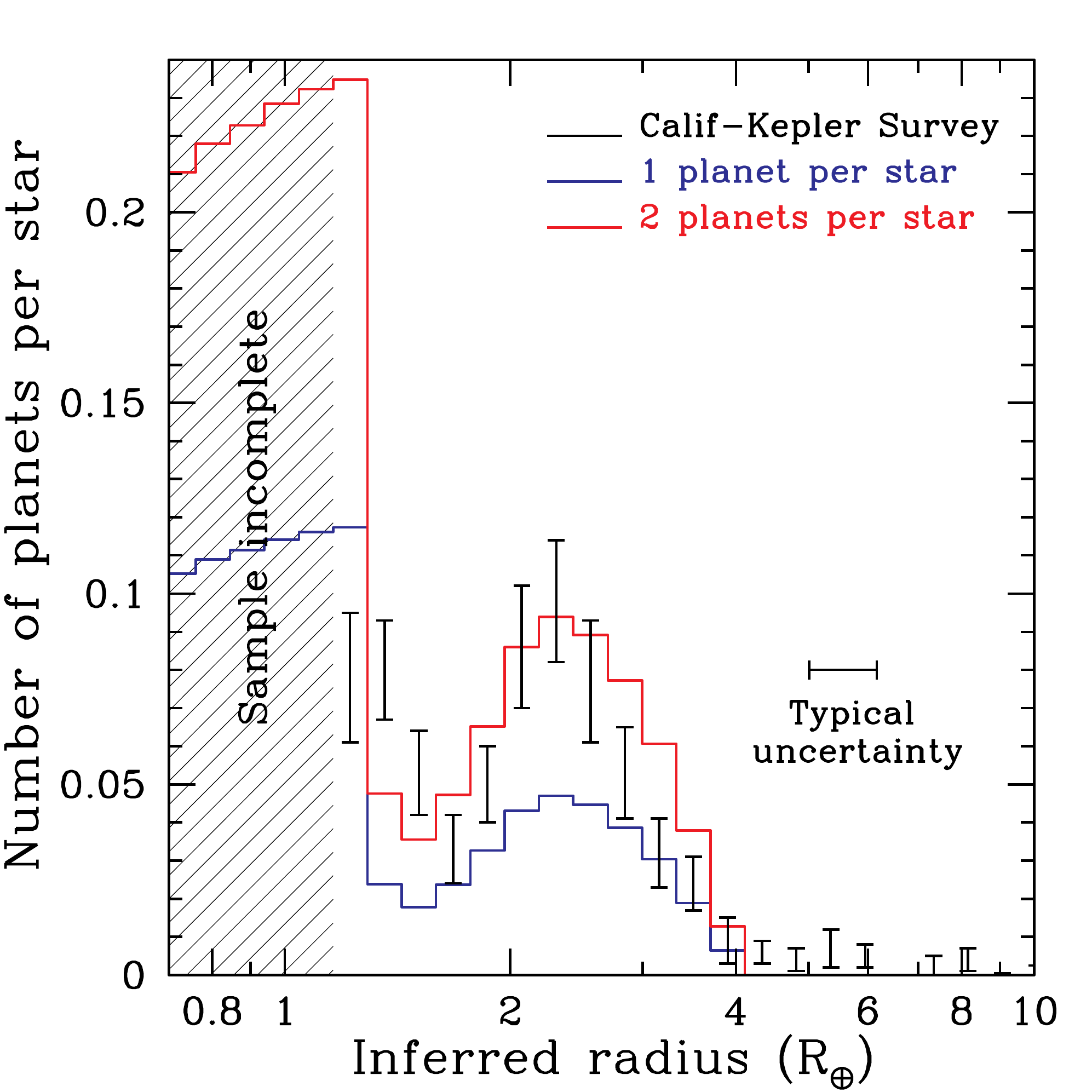}
\caption{Comparison between the radius distribution of small planets from the California-{\it Kepler} Survey to a distribution of planets with rings where 40\% of the planets have rings, with $\theta_{\rm obl}=\pi/2$ and $P(R_p)\propto R_p^{-1}$. Cases with 1 planet per star (blue histogram) and 2 planets per star (red histogram) are considered. Below a radius of $1.16\,R_\oplus$, the sample is incomplete and not plotted.}
\label{fig:histo_compare}
\epsscale{1.0}
\end{figure}

Nevertheless, the exact explanation for the presence of such a H/He envelope and the resulting distribution of radii is still uncertain. It could indeed be due to photoevaporation losses, but there is currently not enough data on exoplanet radii as a function of insolation flux to test this idea in detail. Other mechanisms for getting a small H/He envelope on a rocky core include the delay of gas accretion due to dynamical friction in the protoplanetary disk \citep{Lee14,Lee16}, a secondary atmosphere outgassed during planet formation and evolution \citep{Adams08,Elkins08}, or erosion by impacts that significantly strip large primordial envelopes down to just a few percent of the their initial mass \citep{Liu15,Schlichting15,Inamdar16}.

Although previous theoretical work provides many reasons to think the bimodal radius distribution could be related to the envelope masses, here I consider the speculative idea of whether it could be related to the presence of rings instead. Figure \ref{fig:histogram} demonstrates that a bimodal distribution in radii is expected from the presence of rings, but there are two difficulties. First, to get a sufficiently large secondary peak, the obliquity must be rather large with $\theta_{\rm obl}\approx \pi/2$. This can be somewhat offset with a larger $R_e$ and/or smaller $R_i$ than I have chosen here. Given the uncertainties in these parameters, it is plausible that they are somewhat different that the fiducial values I assume. A second, potentially larger problem is that Figure~\ref{fig:histogram} assumes that all rocky exoplanets have rings, which is almost certainly not the case. Furthermore, as Figure~\ref{fig:histogram3} demonstrates, when rings are only present for a subset of rocky planets, the first peak quickly grows large in comparison to the second. If this is the case, can this model at all be reconciled with the results of \citet{Fulton17}?

To investigate this further, I plot the case when 40\% of rocky exoplanets have rings with $P(R_p)\propto R_p^{-1}$ and $\theta_{\rm obl}=\pi/2$ in Figure \ref{fig:histo_compare}. Note that thus far I have been plotting the {\it fraction of all planets that have a certain radius}, whereas \citet{Fulton17} considers the {\it number of planets per star with a certain radius}. To make the conversion requires understanding the average number of small planets per star \citep{Howard10,Mayor11,Howard12,Dong13,Fressin13,Petigura13,Burke15,Dressing15}, and since this is uncertain, in Figure~\ref{fig:histo_compare} I consider two characteristic values of 1 planet per star (blue histogram, which does not fit the second peak) and 2 planets per star (red histogram, which better represents the data).

This shows that although there are many uncertainties and free parameters that can be varied, in principle, rings can explain the secondary peak in these exoplanet radii. There are main two potential issues with this. First, it requires that on the order of $\sim$40\% of rocky exoplanets need rings. Although this is not 100\%, it still may be uncomfortably large. On the other hand, once a rocky exoplanet has rings, they could last for a consider amount of time (see the discussion in Section~\ref{sec:lifetime}). Second, to get the correct fraction of exoplanets with inferred radii in the range $\approx2-3\,R_\oplus$ implies a much larger fraction of planets per star in the range below $\lesssim 1.2\,R_\oplus$. Currently, radius measurements are incomplete at small radii, but perhaps not at a large enough level to provide the large amount of $\lesssim 1.2\,R_\oplus$ exoplanets needed here.

Alternatively, this work may simply suggest that the sub-Neptunes are subject to contamination by ringed rocky exoplanets, but this may not explain the majority of these objects. These sub-Neptunes therefore should be subject to greater scrutiny to establish there true nature.

\section{Conclusions and Discussion}
\label{sec:conclusion}

Although rocky planets with rings do not exist in our Solar System, it has not been established whether this is true or not for exoplanets. Motivated by theoretical studies that suggest that at least some rocky exoplanets should have rings \citep[e.g.,][]{Hesselbrock17,Barnes02,Piro18}, here I explored what the inferred distribution of radii would be for rocky exoplanets if such rings existed. The main conclusion is that rings lead to an inferred (rather than actual) radius distribution with two peaks, with the first peak being from the intrinsic radius distribution of rocky exoplanets and the second peak from the rings transiting at an average viewing angle. The exact details of this distribution depend on many uncertain factors, such as the underlying radius distribution of rocky exoplanets, the radii of the inner and outer edges of the rings, and the distribution of obliquities of small exoplanets. These are all varied and explored in this work to provide better intuition on how they impact the resulting inferred radius distribution.

Furthermore, I discuss the expected lifetime of these rings assuming that they can form, which is found to be \mbox{$\gtrsim4\times10^8\,{\rm yrs}$.} This makes it at least plausible that the rings can be present for a non-negligible fraction of rocky exoplanets. In detail, these rings may come and go and alternatively be moons instead as the ring undergoes viscous evolution and gravitational instabilities, so more sophisticated time dependent models should be developed to address the duty cycle for these rings in detail.

I also consider the speculative idea of whether the secondary peak in the radii of small planets \citep{Fulton17} can be explained by the presence of rings. This is potentially difficult, but not ruled out. As shown by Figure \ref{fig:histo_compare}, a rather extreme maximum obliquity of $\theta_{\rm obl}\approx\pi/2$ is needed. In addition, to get a sufficient fraction of inferred radii in the range $\approx2-3\,R_\oplus$ requires an even larger fraction of small radii $\lesssim 1.2\,R_\oplus$, which may be difficult to reconcile with exoplanet measurements.

Even if this model cannot completely explain the secondary peak in the range of $\approx2-3\,R_\oplus$, it suggests that rings could be an important contaminate. Therefore, this radius range should be subject to further scrutiny to address the fraction of rocky exoplanets with rings. One way to distinguish between rings and the favored scenario of an inflated H/He envelope would be via transmission spectroscopy, since the transit of rocky rings would be wavelength independent. {In fact, GJ~1214b is a relatively low density exoplanet with a radius of $\approx 2.8\,R_\oplus$, and its transimission spectra is rather featureless \citep{Bean10,Berta12,Kreidberg14}. Unfortunately, it may be difficult to definitively determine if there are rings in this way, since both rocky rings and a cloudy atmosphere would be rather wavelength independent. Indeed a gas envelope is the favored explanation for GJ~1214b \citep[e.g.,][]{Rogers10}.} Nevertheless, evidence for rings around rocky exoplanets would be another amazing way that extrasolar planetary systems differ from our own.

\acknowledgments
I thank Johanna Teske for pointing out the small planet radius gap issue and providing feedback on a previous draft. I thank Tiffany Meshkat for helpful discussions, and Jason Barnes for bringing to my attention the inward migration of Phobos, which helped inspire this work. I also thank N\'{e}stor Espinoza, Matthew Kenworthy, Fran\c{c}oise Roques, Pablo Santos Sanz, Mario Sucerquia, Vincent Van Eylen, and Andrew Youdin for helpful comments and recommendations.

\begin{appendix}

\section{Probability Distributions}

For the Monte Carlo runs, I use a probability density for the planetary radius distribution,
\be
	P(R_p) \propto R_p^{-\eta}.
\ee
which is defined such that
\be
	\int_{R_1}^{R_2} P(R_p) dR_p = 1,
\ee
where $R_1$ and $R_2$ are the smallest and large radii for the distribution of planets, respectively. Here for completeness I describe how this is implemented.

A random number $\Psi$ is chosen to be between $0$ and $1$. From this the planet radius for $\eta=0$ is given as
\be
	R_p = R_1+(R_2-R_1)\Psi,
\ee
for $\eta=1$
\be
	R_p = R_1\exp \left[ \ln \lp \frac{R_2}{R_1}\rp\Psi \right],
\ee
and for $\eta>1$
\be
	R_p = \left[ \frac{1}{R_1^{\eta-1}}
    -\lp \frac{1}{R_1^{\eta-1}}-\frac{1}{R_2^{\eta-1}} \rp \Psi\right]^{-1/(\eta-1)}.
\ee
The random number $\Psi$ is generated using the \texttt{RAN2} routine from \citet{Press92}.

\end{appendix}

\end{document}